\documentclass[useAMS,usenatbib]{mn2e}
\usepackage{txfonts}
\usepackage{longtable}  
\usepackage{rotating}
\usepackage{natbib}
\usepackage{graphicx}
\usepackage{graphics}
\usepackage{psfrag}
\usepackage{amssymb}
\usepackage{multicol}
\usepackage{lscape}
\bibpunct{(}{)}{;}{a}{}{,}
\def\Teff{\ensuremath{T_{\mathrm{eff}}}}
\def\logTeff{\ensuremath{\log T_{\mathrm{eff}}}}
\def\logg{\ensuremath{\log g}}

\def\vsini{\ensuremath{{\upsilon}\sin i}}
\def\kms{$\mathrm{km\,s}^{-1}$}

\def\veq{$V_{\mathrm{eq}}$}

\def\logl{\ensuremath{\log L/L_{\odot}}}

\def\M{\ensuremath{M/M_{\odot}}}

\def\stereo{\textit {STEREO}}
\def\asec{\hbox{"\hskip-3pt\,\,}}
\title[A \stereo\ photometric study of chemically peculiar stars]
{A photometric study of chemically peculiar stars with the \textit{STEREO} 
satellites. II. Non-magnetic chemically peculiar stars\thanks{Data obtained with 
the Heliospheric Imager instruments on board the \stereo\ spacecraft.}} 

\author[E. Paunzen et al.]{E. Paunzen,$^{1,2,5}$\thanks{epaunzen@physics.muni.cz}
			   K.\,T. Wraight,$^{3}$
         L. Fossati,$^{3,4}$
			   M. Netopil,$^{5}$		   
			   G.\,J. White$^{3,6}$ and
			   D. Bewsher$^{7}$\\
$^{1}$Department of Theoretical Physics and Astrophysics, Masaryk University,
Kotlarska 2, CZ-611~37 Brno, Czech Republic\\
$^{2}$Rozhen National Astronomical Observatory, Institute of Astronomy of the Bulgarian 
Academy of Sciences, P.O. Box 136, BG-4700 Smolyan, Bulgaria\\
$^{3}$Department of Physical Sciences, Open University, 
Walton Hall, Milton Keynes MK7 6AA, UK\\
$^{4}$Argelander Institut f{\"u}r Astronomie, Auf dem H{\"u}gel 71, Bonn, 53121, Germany \\
$^{5}$Institut f{\"u}r Astrophysik, Universit{\"a}t Wien, 
T{\"u}rkenschanzstrasse 17, 1180 Wien, Austria\\
$^{6}$Space Science and Technology Department, STFC Rutherford 
Appleton Laboratory, Chilton, Didcot, Oxfordshire, OX11 0QX, UK\\
$^{7}$Jeremiah Horrocks Institute, University of Central Lancashire, Preston, Lancashire, PR1 2HE, UK}
\begin{document}

\date{}

\pagerange{\pageref{firstpage}--\pageref{lastpage}} \pubyear{2012}

\maketitle

\label{firstpage}

\begin{abstract}
We have analysed the photometric data obtained with the \stereo\ spacecraft
for 558 non-magnetic chemically peculiar (CP) stars to search for rotational and pulsational variability. 
Applying the Lomb-Scargle and the phase dispersion minimisation methods, 
we have detected photometric variability for 44 objects from which 35 were previously
unknown. The new objects are all bright stars on the Ecliptic Plane (magnitude range 
$4.7 < V < 11.7$) and will therefore be of great interest to studies of stellar structure 
and evolution.  In particular, several show multiple signals consistent with hybrid 
$\delta$~Scuti and $\gamma$~Doradus pulsation, with different periodicities allowing 
very different regions of the stellar interior to be studied. There are two subgroups of stars in our sample: the cool
metallic line Am (CP1) and the hot HgMn (CP3) stars. These objects fall well inside 
the classical instability strip where $\delta$ Scuti, $\gamma$ Doradus and 
slowly pulsating B-type stars are located. We also expect to find periods
correlated to the orbital period for CP1 objects as they are mostly members of
binary systems. For CP3 stars, rotationally-induced variability is still a matter
of debate. Although surface spots were detected, they are believed to produce only
marginal photometric amplitudes. So, periods from several
hours to a few days were expected for these two star groups. The \stereo/HI-1 data are well matched to studies of this frequency domain, 
owing to the cadence of approximately 40 minutes, and multiple epochs over four 
and a half years. The remaining 514 stars are likely to 
be constant in the investigated range from 0.1 to 10\,days. In some cases, the presence 
of blending or systematic effects 
prevented us from detecting any reliable variability and in those cases we 
classified the star as constant. We discuss our results in comparison to 
already published ones and find a very good agreement. Finally, we have calibrated
the variable stars in terms of the effective temperature and luminosity in order
to estimate masses and ages. For this purpose we used specifically developed
calibrations for CP stars and, when available, HIPPARCOS parallaxes. All but two
objects cover the stellar mass range from 1.5 to 5~M$_{\sun}$ and are 
located between the Zero- and Terminal Age Main Sequence.
\end{abstract}

\section{Introduction}\label{introduction}
In the first paper of this series \citep[][WFN12 hereafter]{wraight2012}, 
we presented photometric time series of 337 magnetic chemically peculiar 
stars from the NASA's twin \stereo\ spacecraft. In total, 82 objects were
identified as variable caused by rotation in the presence of surface spots
\citep{mikulaek2011}.

As a further step, the light curves of non-magnetic chemically
peculiar (CP) stars of the upper main sequence were analysed. The sample consists of cool
metallic line Am (CP1) and the hot HgMn (CP3) stars. Most CP1 and CP3 stars are members of a 
binary system in which the rotation of the stars has been slowed down by tidal interaction.
The members of these groups have 
stellar masses from 1.5 to 5~M$_{\sun}$ and are located between the zero- and
terminal age main sequence. Thus, they fall well inside the classical instability 
strip where $\delta$ Scuti, $\gamma$ Doradus and slowly pulsating B-type \textbf{(SPB)}
stars are found. Indeed, these types of pulsators were found among CP1
and CP3 stars \citep{alecian09, smalley11}. 
The \stereo/HI-1 data have a cadence 
of approximately 40 minutes, and multiple epochs over four 
and a half years. Therefore, these data
sets are perfectly suited to find the predicted periods of variability.

We have analysed photometric data, obtained with the \stereo\ spacecraft,
of 558 stars listed in the \citet{renson09} catalogue as known or suspected 
CP1/3 objects and identified 44 variable stars from which
35 were previously unknown. This catalogue is the most comprehensive available, although the collected observations 
are rather inhomogenous and little is known of many of the stars in our sample.  It is therefore important to 
constrain which stars are constant, within our detection limits, in order to focus immediate future efforts on 
the variable stars.  
Therefore, although 399 stars were clearly constant or the data of insufficient quality or 
quantity for analysis, 115 stars are also listed that were examined in as much detail as the 44 variable stars.  
In some cases weak signals were observed at low statistical significance, or systematic effects may have prevented 
the detection of low amplitude variability, however many appear also to be constant within the 
sensitivity of \stereo/HI-1. These stars may be constant, or variable with an amplitude below the 
sensitivity of \stereo/HI-1, variable with a periodicity outside the range 0.1 to 10\,days, variable in a part of 
the spectrum which \stereo/HI-1 is not sensitive to or rotational variables seen pole-on.
\section{Observations and data analysis}\label{obs_data}
In the following we give a short overview of the observations
and the applied data analysis. This should serve as a guideline for the
understanding of the apparent limitations of the result. For further details the reader is referred to the description in WFN12.

The observations were obtained by the twin \stereo\ spacecraft using the Heliospheric Imager cameras (\stereo/HI-1A and \stereo/HI-1B). 

Each image has a field of view of 20 degrees by 20 degrees
with a pixel scale of 70\,\asec\ per pixel, centred 14 degrees away from the centre of the Sun. A single filter, with a spectral 
response mostly between 6300\,\AA\ and 7300\,\AA, is used. The integration time was 40 seconds with a summed image cadence of 40 minutes. In general, an object
remains in the field of view of the \stereo/HI-1A/B imagers for 19/22\,days. 
In summary, each data set covers a total span of four and a half years, with about twenty days continuous observations per HI camera, and gaps of about one year.

The light curves were cleaned by removing all data points more than four standard 
deviations away from the weighted mean magnitude. Polynomial detrending is then carried 
out using a 4th order polynomial in order to correct for any existing residual trends.

The time series analysis was performed in several stages. First, 
synthetic light curves are constructed and the least-squared error of the model 
compared to the actual light curve is measured. Then the period and amplitude
was iteratively determined. The periodograms and light curves 
were visually inspected, primarily to extract from the sample the objects 
which appeared clearly constant.
Additionally, we classified as constant those stars which were so faint that 
any signal would be likely due to noise or if systematic effects were so 
extreme that the data were unusable. The same classification was given to the 
stars for which the lack of data prevented the reliable detection of any 
variability. The list of those 399 stars is given in Table~\ref{tab:junk}. 

The final stages of the detailed analysis were done with
Peranso\footnote{\tt http://www.peranso.com}. Two algorithms were applied, to 
cross-check each other and avoid duplicating weaknesses. In each case, we 
searched for periods between 0.1 and 10\,days, 
although in a few individual cases a search was made outside this range. The 
Lomb-Scargle method \citep{scargle82} was employed in the period domain, whereas 
the phase dispersion minimisation (PDM) method \citep{stellingwerf78} was 
employed in the frequency domain. The latter are using $\Theta$, defined in 
\citet{stellingwerf78} which gives a direct indication of the significance 
of a certain period. We then examined the significant features 
in the periodogram produced with each method to extract the most likely 
period, its uncertainty and the epoch of the first maximum in the \stereo\ 
light curve. 
\section{Results}\label{results}
As a result, we obtained:
\begin{itemize}
  \item 399 stars without a reliable detection (Table~\ref{tab:junk})
	\item 115 constant stars (Table~\ref{tab:const})
	\item 44 variable stars (Table~\ref{tab:variable}) 
\end{itemize}

As mentioned in Sect.~\ref{obs_data}, Table~\ref{tab:junk} lists all stars that
were immediately classified as constant, but it also includes stars for which 
the photometry was clearly affected by systematic effects and/or by blending, 
making the detection of any periodicity impossible. Table~\ref{tab:const}, on the
other hand, includes all stars for which our
analysis did not identify the presence of variability.

Each Table lists 
\begin{itemize}
  \item Column 1: star name
	\item Column 2: identification number by \citet{renson09} 
	\item Column 3 and 4: equatorial coordinates in degrees
	\item Column 5: average $V$ magnitude
	\item Column 6: spectral classification and chemical peculiarity given by \citet{renson09}
	\item Column 7: CP classification
\end{itemize}

The derived CP classification is a combination of the Catalogue of Ap, HgMn and Am stars 
by \citet{renson09} as well as the extensive list of spectral types by \citet{skiff2012}.
If contradicting classifications within these references were found, a question mark was set. 

Table~\ref{tab:variable} lists the variable objects 
\begin{itemize}
  \item Column 8 and 9: genuine period and its uncertainty
	\item Column 10: MJD of the epoch for the first recorded maximum 
	\item Column 11: flag indicating the possible presence of blending (B), systematic (S) effects,
	                 or an exceptionally strong signal (``*") (see WFN12, for more details)	                 
	\item Column 12: period found in the literature with reference
\end{itemize}

For each star in this table, we generated a classical periodogram using 
the PDM method (Fig.~\ref{fig:allperiodograms}) and a phase-folded light curve
(Fig.~\ref{fig:alllightcurves}). We use $\Theta$ as a direct indication of the 
periods' significance.

\begin{figure}
\begin{center}
\includegraphics[width=85mm,clip]{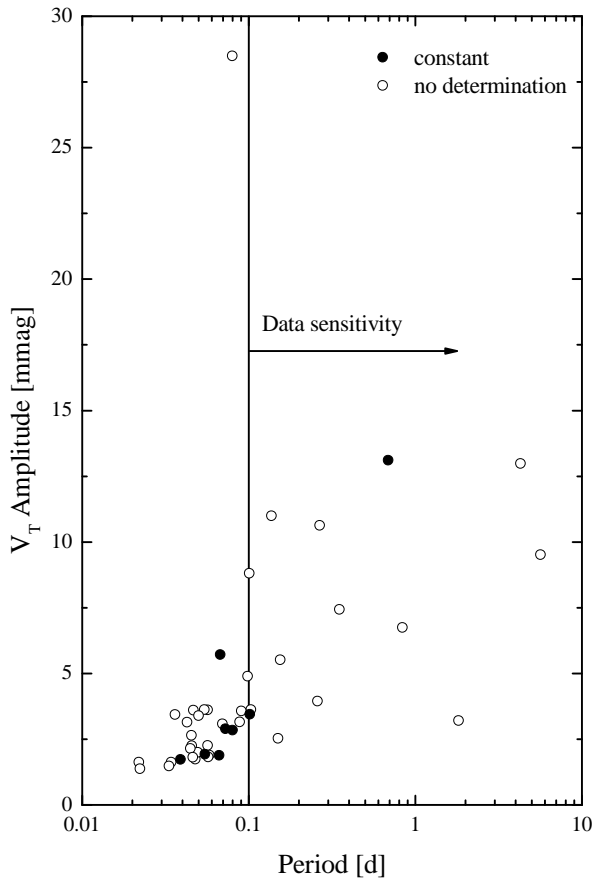}
\caption{Periods and amplitudes of 44 variables identified by \citet{smalley11} 
for which no significant frequency was found in this work.}
\label{fig:smalley} 
\end{center} 
\end{figure}
\begin{figure}
\begin{center}
\includegraphics[width=85mm,clip]{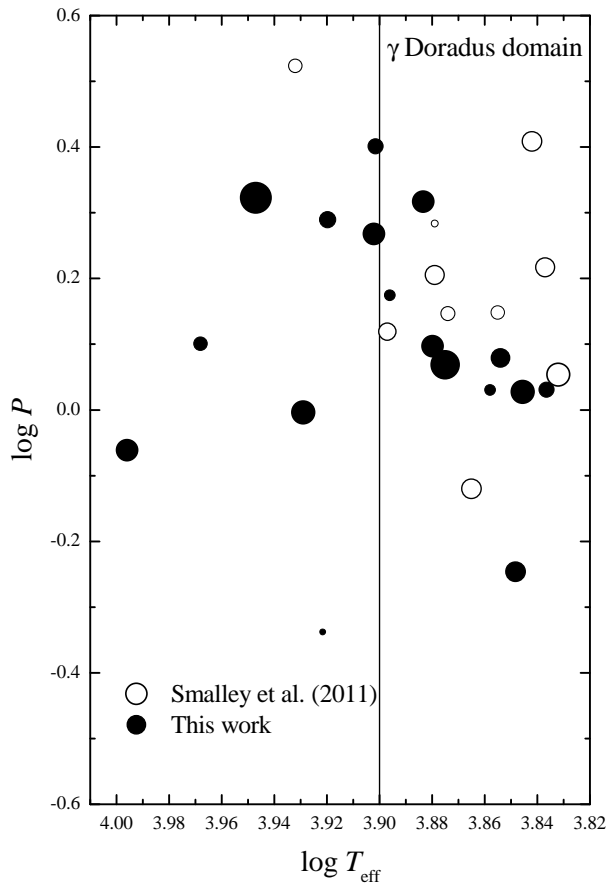}
\caption{The $\gamma$ Doradus stars taken from \citet{smalley11}
and our variables in the temperature - period domain.
The symbol sizes are proportional to the amplitudes. Objects which are
hotter than \logTeff$>$\,3.9 are not pulsators but show variability
due to multiplicity.}
\label{fig:gammador} 
\end{center} 
\end{figure}
\begin{figure}
\begin{center}
\includegraphics[width=85mm,clip]{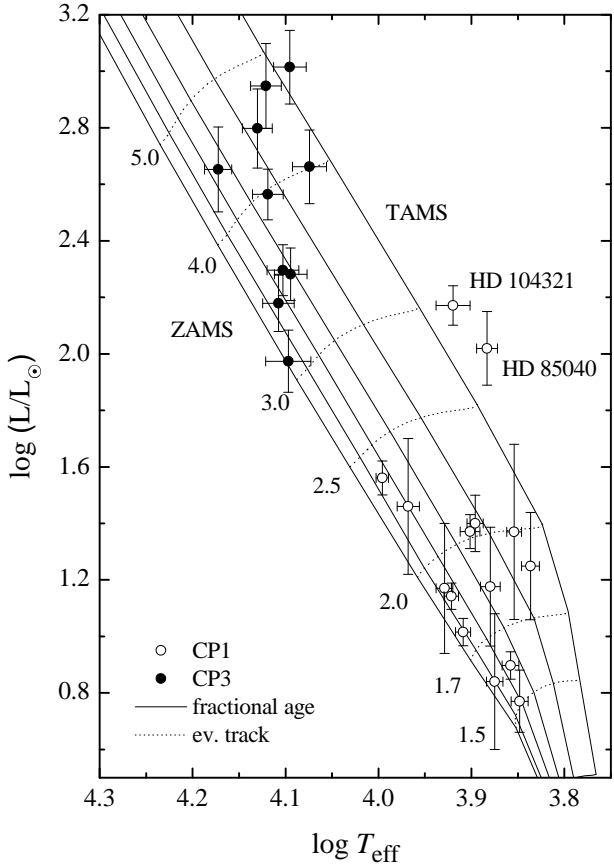}
\caption{Hertzsprung-Russell diagram for the CP1 (open circles) and CP3 (filled
circles) stars listed in Table~\ref{tab:variable} and for which we
derived both \Teff\ and \logl. The solid lines represent the lines of equal
fractional age. The dashed
lines are the main sequence evolutionary tracks for solar metallicity \citep{schall92} used to derive the 
stellar masses. Two objects, HD~85040 and HD~104321, lie above the Terminal Age Main Sequence (TAMS).}
\label{fig:hr} 
\end{center} 
\end{figure}

\section{Discussion}

Recently, \citet{smalley11} presented SuperWASP photometric data of
1600 CP1 stars at a precision level of 1\,mmag. They found about 200 objects which
where identified as pulsating $\delta$ Scuti, $\gamma$ Doradus and/or hybrids. 
Their target selection criteria was substantially similar to ours. We cross-matched the two samples and
identified 36 objects for which we were not able to derive any variability, 8 stars, which we classified as constant,
and one common variable.
We further investigated the reason for the apparent inconsistency. Figure~\ref{fig:smalley} shows the
periods versus the amplitudes for those 44 objects. Most of these stars have periods
and amplitudes below our instrumental sensitivity.
However, we also looked for very short period variability by extending the period search 
down to 0.064\,d when using the Lomb-Scargle method and further, to 18/d, when using 
Phase Dispersion Minimisation. As a result, our periodograms are mostly featureless
not showing any significance. 

In the context of the CP1 stars we analysed, the signal for TYC 1876-325-1 (0.1038\,d) is 
the only one where the weak signal found seems likely to have originated in the target star, 
while the weak signals for HD\,243093 (2.8426\,d) and HD\,146053 (1.2202\,d) seem more likely to have 
originated in a neighouring star, although neither of these two stars has a variable star 
recorded in their vicinity. If we compare the signals extracted for these two stars with those 
recorded as such in WFN12, only the signal from HD\,146053 might be strong enough, but the likelihood 
of blending discouraged us from reporting it.

Figure~\ref{fig:gammador} shows the $\gamma$ Doradus stars published by \citet{smalley11} and our 
variables in the temperature - period domain.
As a temperature limit for the $\gamma$ Doradus domain, we adopted \logTeff$<$\,3.9 which also includes
all types of hybrid pulsators \citep{balona11}. Stars hotter than that are most certainly not members
of this variable star group. The amplitudes of the $\gamma$ Doradus type pulsation in that paper
(Figure 6 therein) are found to be up to 25\,mmag. The symbol sizes are proportional to the amplitudes which
are all below 25\,mmag.

In the following, we discuss individual stars from Table~\ref{tab:variable} which deserve special attention.

{\it AAO+23\,222, AAO+25\,242, and AAO+28\,487:} The found amplitudes for these stars are between
30 and 41\,mmag, respectively. Due to the lack of photometric measurements, we were not able to calibrate
the effective temperatures for these objects. Even if they fall into the domain of $\gamma$ Doradus
pulsators (Fig.~\ref{fig:gammador}), their amplitudes are too large \citep{balona11}.

{\it HD~23607:} \citet{fox2006} analysed the pulsational characteristics of this Pleiades
member in more details. In 1997, a STEPHI campaign was dedicated to this star. Up to now, 
no long term variations have been reported in the literature. Our deduced period of 1.8526\,d
is compatible with an orbit due to undetected multiplicity or $\gamma$ Doradus pulsation.
However, the periodogram shows numerous frequencies shorter than 1\,d supporting hybrid
$\delta$ Scuti - $\gamma$ Doradus characteristics. HD~23607 lies just on the edge 
(Fig.~\ref{fig:gammador}) where these objects can be found.

{\it HD~23950:} This star has the highest \vsini\ of all objects among our variable
sample with 82.5$\pm$3.8\,\kms. \citet{winzer1974} published a period of 1.1\,d which 
was flagged with a question mark. Our time series analysis results in a period of 3.2509\,d.
As described in WFN12, we calculated the equatorial velocity (\veq) from the formula of the oblique 
rotator model on the basis of \Teff\ and \logl. Taking the above mentioned periods, we get 
\veq\ of $\sim$95 and $\sim$32\,\kms, respectively (Fig.~\ref{fig:vsini}). In contrary to the shorter period, the
longer period is clearly not compatible with the oblique rotator model and could be interpreted
as a SPB characteristics.

{\it HD~27628:} A well known $\delta$ Scuti type pulsator (V775~Tau) and member of
the Hyades \citep{perryman98}. The orbital period published by \citet{griffin2012}
is twice the variability we derived from the \stereo\ data.

{\it HD~31592, HD~39078, and HD~104321:} These stars have periods between approximately 
one and two days, but are too hot for being $\gamma$ Doradus pulsators.

{\it HD~42066 and HD~56152:} For each star, we detected two clear frequencies in the
domain of the $\gamma$ Doradus pulsators. However, both stars are too hot (Fig.~\ref{fig:gammador}),
and the amplitude for HD~42066 is too large for such kind of variability. This might be a sign of
long-period pulsations in close binaries which are tidally excited \citep{handler02}. These two
stars are, therefore, very interesting targets for spectroscopic follow-up observations in order
to get radial velocity data.

{\it HD~49606:} 
We find a period of 2.2661\,d whereas in the literature values of 1.10503 and 3.35\,d
are published. From our periodogram, we conclude that the shorter period 
does not seem to be present, but there are several peaks {around 3.3\,d, although they
are less significant than the ones listed in Table~\ref{tab:variable}.

{\it HD~122911, HD~144844, HD~202671, and HD~211838:} For these objects we found
at least one frequency typical for $\delta$ Scuti and one for $\gamma$ Doradus
type pulsation. Therefore, these objects might be hybrid pulsators.

{\it HD~138124:} With two detected clear frequencies, 
we conclude that this object is a $\gamma$ Doradus pulsator.

{\it HD~244698:} \citet{smalley11} published two $\delta$ Scuti like frequencies 
with amplitudes of 1.35 and 1.69\,mmag, respectively. Its periodogram shows also
some very low amplitude frequency in the range where we find a significant period
of 0.9993\,d. However, the $\Theta$ value is just on the detection limit.

In order to locate the variable stars from Table~\ref{tab:variable} in the Hertzsprung-Russell 
(HR) diagram, we compiled Johnson $UBV$, Str{\"o}mgren, and Geneva photometry from the General 
Catalogue of Photometric data \citep{mermilliod97}.
These data were used to derive the effective temperature on the basis of the calibrations 
given by \citet{netopil08}. The final \Teff, listed 
in Table~\ref{tab:variable}, is the average \Teff\ obtained 
calibrating the different colors, while the standard deviation and the number 
of averaged temperature values are given in parentheses. When available, we adopted 
the spectroscopic \Teff\ listed in \citet{netopil08}, and these 
cases are indicated with a ``99" instead of the number of averaged 
temperatures. Similarly to WFN12 also a spectral energy distribution (SED) fitting was 
performed in order to obtain an additional  
temperature estimate. However, no correction was applied since CP1/3 objects do not show anomalous 
colours. The stars, for which the latter method was the only possibility to deduce reddening, are 
flagged with ``50" in the \Teff\ column of Table~\ref{tab:variable} (see WFN12 for details).

If available, parallax from the HIPPARCOS catalogue \citep{leeuwen} were
used to determine the luminosity (\logl) on the 
basis of the Johnson $V$ magnitude and the interstellar reddening $E(B-V)$ determined via 
the different photometric systems or 
via SED fitting, using a total-to-selective absorption ratio of $R=3.1$. Since according to \citet{netopil08} 
the bolometric correction for CP1/3 stars do not differ from normal ones, we used the tabulated corrections listed by 
\citet{flo96}. 
As next step, we determined the stellar mass and fractional age 
($\tau$ - fraction of main sequence lifetime completed) using the 
evolutionary tracks for solar metallicity given by \citet{schall92}. 

Figure~\ref{fig:hr} shows the final HR diagram for all stars listed 
in Table~\ref{tab:variable} and for which we derived both \Teff\ and \logl.
The two CP groups are clearly distinct according to their \Teff. All,
but two objects, HD~85040 and HD~104321 lie well between the ZAMS and TAMS. 
We have investigated the two outliers in more details.

{\it HD~85040:} \citet{fremat2005} analysed this spectroscopic triple system
using a disentangling technique to extract the individual contributions of the three components
to the composite spectrum. The inner binary consists of two Am components, whereas 
the distant third component is confirmed to be a $\delta$ Scuti star with normal 
chemical composition. We report the detection of three periods of which two
(0.0834 and 0.0881\,d) are from the pulsational component. The third one (2.0735\,d)
is half the orbital period. \citet{fremat2005} estimated that all three components
have almost the same effective temperature of 7500\,K and equal luminosities. Taking
this into account in our HR diagram, this object shifts down to the two solar
mass evolutionary track.

{\it HD~104321:} This spectroscopic binary system has a mass ratio of about 0.47,
$m_{\mathrm{1}}$\,=\,2.2\,M$_{\sun}$ and an orbital period of about 283\,d \citep{ducati2011}.
\citet{shorlin2002} derived a photometric \logg\ of 3.51 which is compatible with our estimates.
Thus, this system seems to be located close to the TAMS. In addition, we checked the SED
with the tool by \citet{robitaille07}. Figure~\ref{fig:hd104321} shows the 
flux distribution according to the photometric measurements overplotted with the
corresponding stellar atmosphere model.
The best fitted effective temperature of 8000\,K is
in perfect agreement with the estimates from all different photometric systems.
At 4.6\,$\mu$m, a bump exceeding 3$\sigma$ of the listed photometric error, 
\citep{wright10} can be found which might be due to the companion.

As a test for the oblique rotator model to explain the periods of the CP3 stars, we 
calculated the equatorial velocities \veq\ from the formula published by \citet{preston71}:
\begin{equation}
V_{\mathrm{eq}} = \frac{50.6 R}{P}
\end{equation}
where $R$ is the stellar radius in solar units and $P$ is the observed period in days. 
Since only the projected rotational velocity can be determined, all stars should fall 
below the given relation assuming a certain stellar radius. Besides HD~23950 (discussed
above), all stars meet the specifications of the oblique rotator model.
\begin{figure}
\begin{center}
\includegraphics[width=85mm,clip]{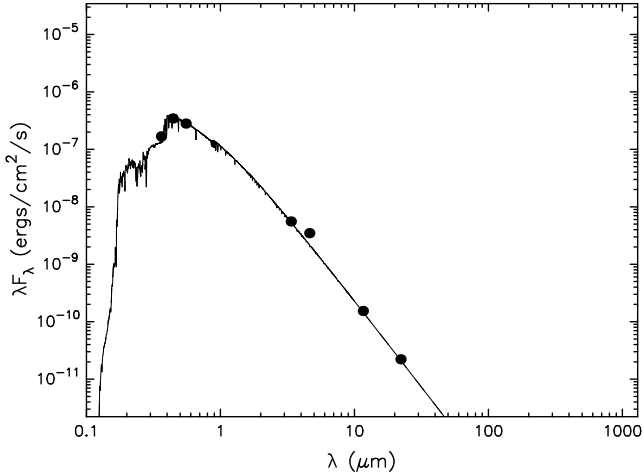}
\caption{The spectral energy distribution of HD~104321 shows a
bump at 4.6\,$\mu$m larger than 3$\sigma$ of the listed photometric error 
\citep{wright10} which could be due to a cool companion.}
\label{fig:hd104321} 
\end{center} 
\end{figure}
\begin{figure}
\begin{center}
\includegraphics[width=85mm,clip]{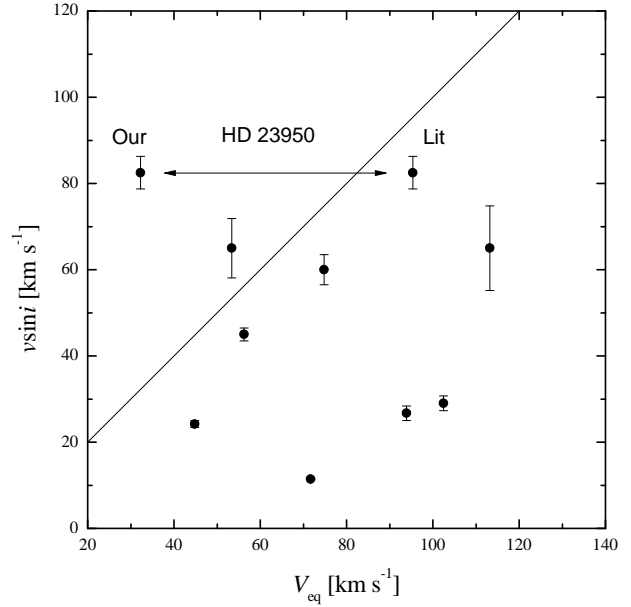}
\caption{Comparison between the observed \vsini\ and the computed \veq\ for 
the stars listed in Table~\ref{tab:variable}. The continuous line is the 
one-to-one relationship. According to the oblique rotator model, all stars
should fall below the given relation. For HD~23950 we plotted the value using
our and the period from the literature.}
\label{fig:vsini} 
\end{center} 
\end{figure}
%
\section{Conclusion}
We present a detailed time series analysis of photometric data for 558 
CP1/3 stars obtained with the \stereo\ spacecraft. 44 stars were found to be variable, 35 of which were 
not previously known as such. The new variables range in brightness $4.7 < V < 11.7$ and are therefore 
particularly interesting for follow-up observations. Several show multiple signals indicative of hybrid 
$\delta$~Scuti and $\gamma$~Doradus pulsation. These bright variable stars will allow for more detailed study 
of stellar structure and evolution. Of particular interest are the variable members of open clusters
(e.g. HD~23607) where age, metallicity, luminosity and therefore mass can be more tightly constrained. 

The period range of the variability we detected among the CP1 stars is compatible with that
expected for $\delta$ Scuti and $\gamma$ Doradus pulsators. In addition, some
already known orbital periods were also photometrically detected.

For the CP3 stars, the periodicity of slowly pulsating B-type stars is in the
same range as that of the known binary systems, therefore, a decision about the
type of variability was not possible, and further observations are needed.
However, \citet{renson01} list 16 variable CP3 stars. With our sample of
7 new ones, we significantly increase the number of known variables of this
subgroup.

The identification of constant stars, within the sensitivity of \stereo/HI-1, is also useful for further analysis,
as these stars might be constant, or variable with an amplitude below the sensitivity of \stereo/HI-1, or 
variable with a periodicity outside the 0.1-10\,days range, or variable in a part of the spectrum that the 
\stereo/HI-1 is not sensitive to, or rotational variables seen pole-on.  It is important to constrain the nature 
of constant CP1/CP3 stars to support further investigations, e.g. suggesting that future observations may require 
higher cadence to observe suspected $\delta$~Scuti variability, or that observations must take place over longer 
continuous periods to detect long-period rotational variability.

The presented sample of constant and variable CP1/3 stars could serve not
only for detailed follow-up observations, but also for the analysis of 
time dependent changes of pulsational and rotational periods \citep{breger2009}. 
Long time basis and continuous observations over many decades are very
much needed to shed more light on pulsational characteristics of these 
objects \citep{mikulaek2011}.
Such bright variable stars are of great value in advancing studies of stellar structure and evolution, as 
these can be studied in great detail with little observational effort. 

\section*{Acknowledgments}
The Heliospheric Imager (HI) instrument was developed by a collaboration 
that included the Rutherford Appleton Laboratory and the University of 
Birmingham, both in the United Kingdom, and the Centre Spatial de Li\'ege 
(CSL), Belgium, and the US Naval Research Laboratory (NRL), Washington DC, 
USA. The \textit{STEREO}/SECCHI project is an international consortium 
of the Naval Research Laboratory (USA), Lockheed Martin Solar and 
Astrophysics Lab (USA), NASA Goddard Space Flight Center (USA), Rutherford 
Appleton Laboratory (UK), University of Birmingham (UK), 
Max-Planck-Institut f\"{u}r Sonnensystemforschung (Germany), Centre 
Spatial de Li\'ege (Belgium), Institut d'Optique Th\'eorique et 
Appliqu\'ee (France) and Institut d'Astrophysique Spatiale (France). 
This research has made use of the \textsc{Simbad} database, operated 
at CDS, Strasbourg, France. This research has made use of version 2.31 
\textsc{Peranso} light curve and period analysis software, maintained at 
CBA, Belgium Observatory http://www.cbabelgium.com.
KTW acknowledges support from a STFC studentship.
This work was supported by grant GA \v{C}R P209/12/0217, 
the financial contributions of the Austrian Agency for International 
Cooperation in Education and Research (CZ-10/2012), and the
Austrian Research Fund via the project FWF P22691-N16.
We thank M.~Hareter and O.~Kochukhov for valuable discussions.
%

\begin{scriptsize}
\onecolumn
\begin{longtable}{lrrrcrcc}
\caption{Basic properties of the CP1 and CP3 stars identified as constant or for which the quality of the data
prevented the detection of any variability.}
\protect\label{tab:junk}\\
\hline\hline \\
Name & \# R09 & RA  & DEC & V  & Spectral & CP    & \vsini \\
     &        & deg.& deg.& mag& Type     & class & \kms   \\
\hline
\endfirsthead
\caption{continued.}\\
\hline\hline \\       
Name & \# R09 & RA  & DEC & V  & Spectral & CP    & \vsini \\
     &        & deg.& deg.& mag& Type     & class & \kms   \\
\hline
\endhead
\hline
\endfoot
\hline
\hline
\endlastfoot
HD\,294	     &    20	&   1.91666 &	 1.09421 &  8.25 &	   A0,NA & CP1   &               \\
HD\,1169	     &   230	&   4.02257 &	 8.11554 &  7.60 &	A8-F2,NA & CP1   &               \\
HD\,3644	     &   985	&   9.78988 &	-3.99267 &  8.91 &	A0-A7,NA & CP1   &               \\
HD\,4570	     &  1230	&  11.91140 &	-4.49765 &  9.34 &	A2-F1,NA & CP1   &               \\
HD\,11386	     &  2850	&  28.01060 &	10.81060 &  7.70 &	A7-F2,NA & CP1   &               \\
HD\,11636	     &  2970	&  28.66000 &	20.80800 &  2.64 &	   A4,NA & CP1   & 71.6(4.0/4)   \\
HD\,12869	     &  3280	&  31.64140 &	22.64830 &  5.03 &	A1-A7,NA & CP1   & 15.0(7.1/2)   \\
HD\,13248	     &  3440	&  32.44400 &	13.17580 &  7.72 &	   A0,NA & CP1   &               \\
\hline                                                  
\end{longtable}%

\end{scriptsize}
\begin{landscape}
\begin{table}
\tiny
\begin{center}
\caption{\label{tab:variable} Properties of the CP1 and CP3 stars identified as photometrically variable.}
\begin{tabular}{lrrrcrccccrrccccc}
\hline
\hline       
Name & \# R09 & RA  & DEC & V  & Spectral & CP    & Period & $\sigma_{period}$ & MJD     & Analysis & Literature & \vsini & \Teff & \logl & \M    & $\tau$ \\
     &        & deg.& deg.& mag& Type     & class & days   & days              & maximum & remarks  & period     & \kms   & K     & dex   &       &        \\
\hline
\hline
HD\,2523            & 600      &   7.28586 &   11.32020 &  8.10 &      A8-F2,NA & CP1   &   0.5668 &   0.0001 & 54167.607 &    B &                                       &               & 7050(-/1)     & 0.77(0.11)   & 1.51(0.06) & 0.37(0.29) \\
HD\,7374            & 1850     &  18.53180 &   16.13350 &  5.97 &   B9,Si,Hg,Mn & CP3   &   2.8240 &   0.0019 & 54187.360 &    B &C98:2.8                                & 24.2(0.8/4)   & 12811(68/4)   & 2.18(0.10)   & 3.40(0.15) & 0.31(0.25) \\
HD\,21437           & 5350     &  52.00580 &   20.46430 &  6.89 &      A3-F3,NA & CP1   &   0.7981 &   0.0002 & 54208.748 &    B &                                       &               &               &              &            &            \\
HD\,23607           & 6050     &  56.83060 &   24.13910 &  8.28 &      A4-F2,NA & CP1   &   1.8526 &   0.0007 & 54213.398 &   BS &S09+W:0.047                            & 10.0(1.5/1)   & 7981(91/3)    &              & 2.16(0.45) &            \\
HD\,23950           & 6130     &  57.47950 &   22.24410 &  6.07 &   B9,Si,Hg,Mn & CP3   &   3.2509 &   0.0020 & 54216.059 &    B &C98:1.1                                & 82.5(3.8/4)   & 12504(245/4)  & 1.97(0.11)   & 3.11(0.17) & 0.03(0.18) \\
HD\,27628           & 7080     &  65.51470 &   14.07720 &  5.72 &      A3-F2,NA & CP1   &   1.0718 &   0.0003 & 54220.184 &    S &S09+W:0.0625                           & 29.5(5.0/4)   & 7210(90/99)   & 0.90(0.05)   & 1.60(0.04) & 0.49(0.13) \\
\hline                                                  
\end{tabular}
\end{center}
\end{table}
\smallskip

\noindent
B: \citet{bychkov2005}\\
C98: \citet{catalano1998}\\ 
C01: \citet{renson01}\\
S09: \citet{samus09}\\
S11: \citet{smalley11}\\
W: \citet{watson2006}
\end{landscape}

\begin{center}
\begin{figure*}
\includegraphics[width=165mm,clip]{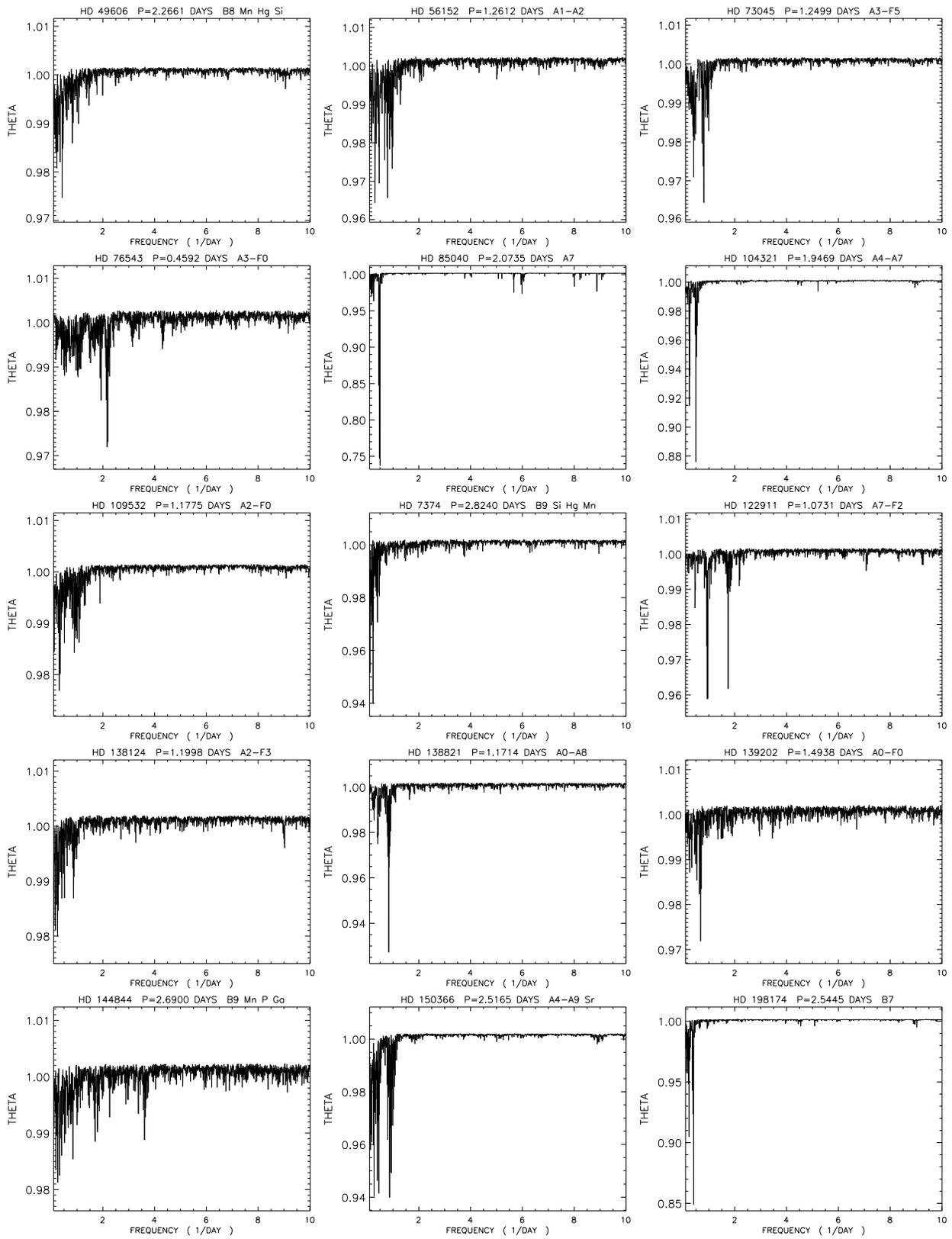}
\caption{Periodograms obtained for the CP1/3 stars, listed in
Table~\ref{tab:variable}.}
\label{fig:allperiodograms} 
\end{figure*}
\begin{figure*}
\includegraphics[width=165mm,clip]{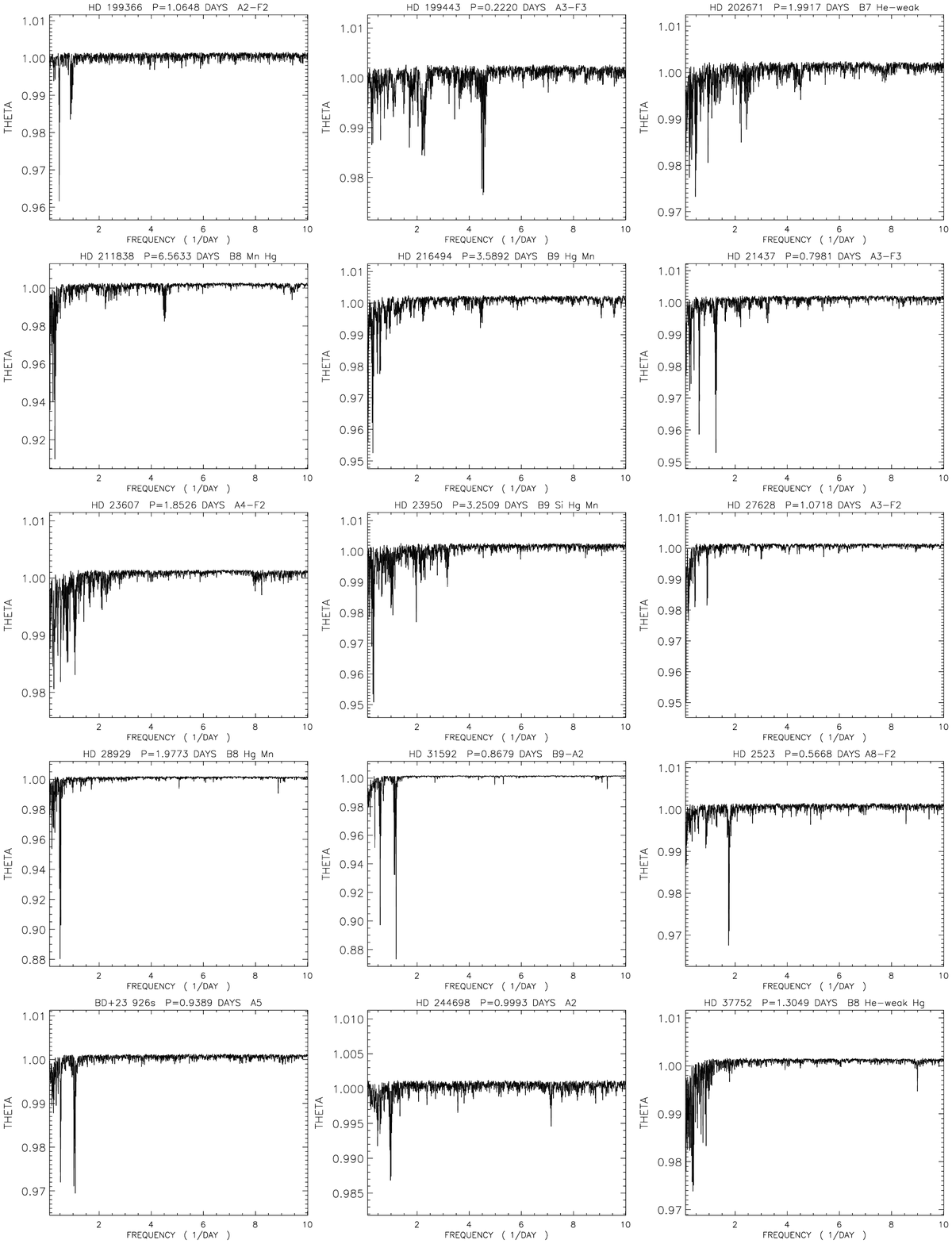}
\end{figure*}
\begin{figure*}
\includegraphics[width=165mm,clip]{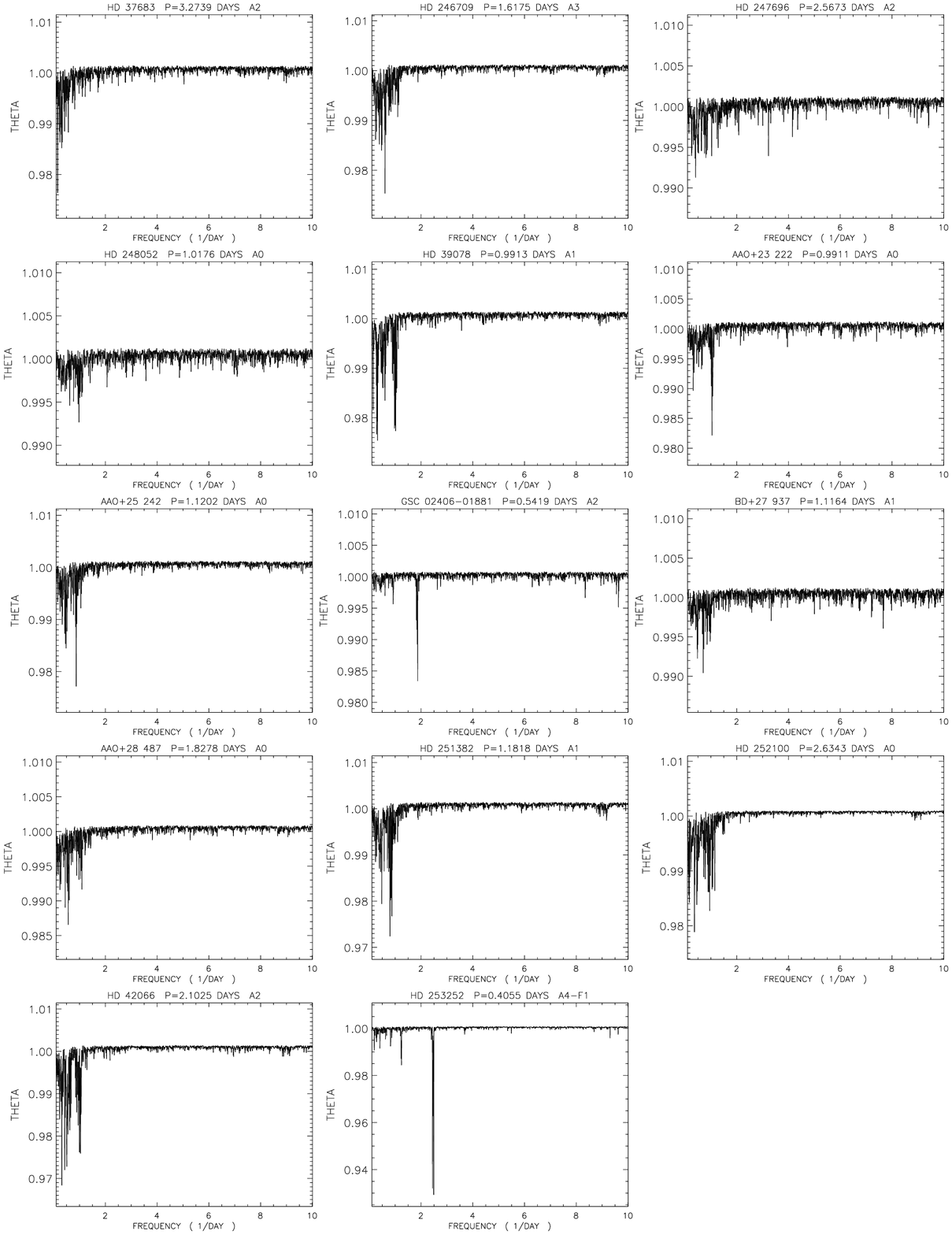}
\end{figure*}
\end{center} 
\begin{center}
\begin{figure*}
\includegraphics[width=165mm,clip]{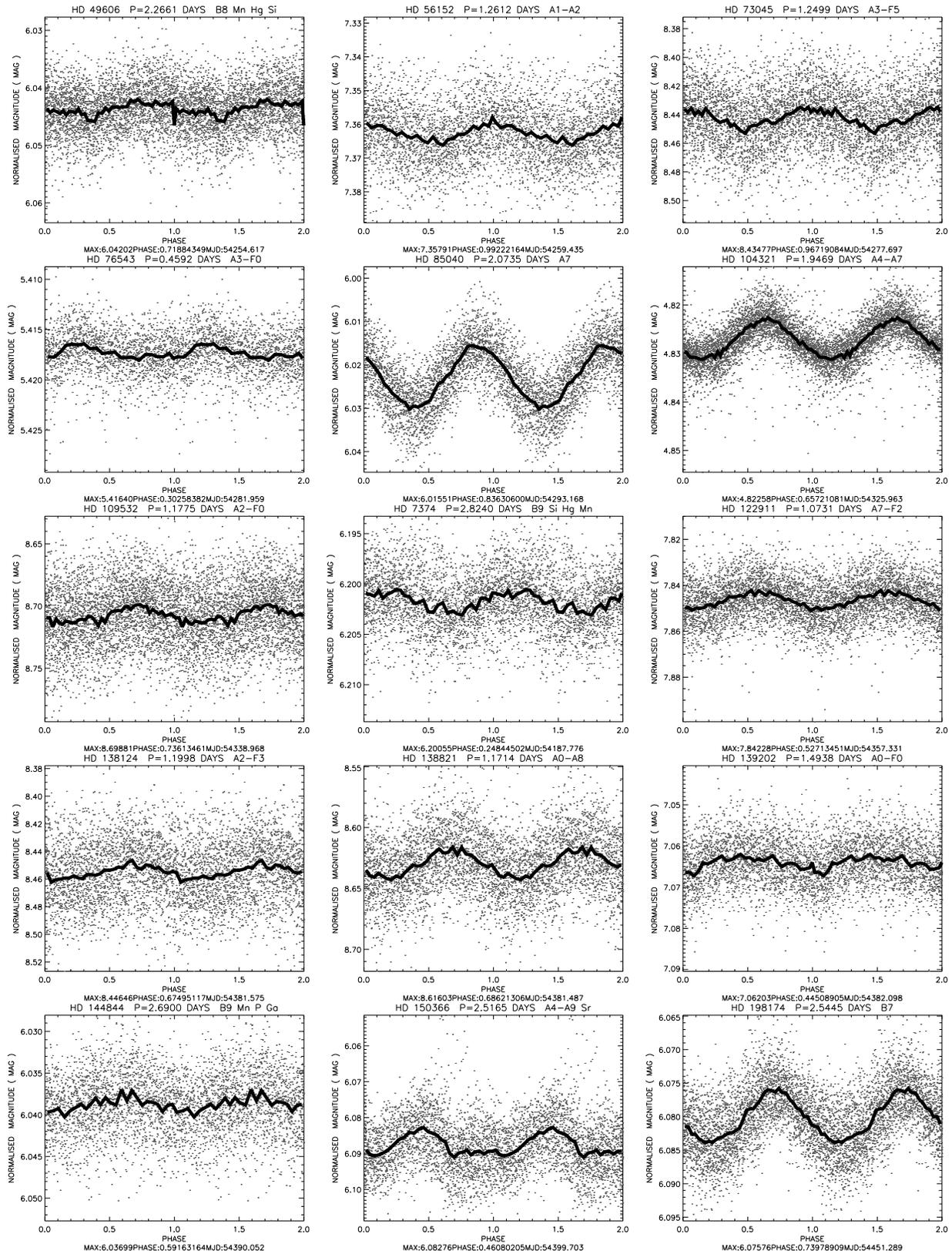}
\caption{Light curves obtained for the CP1/3 stars, listed in
Table~\ref{tab:variable}, phase-folded on the most significant period.}
\label{fig:alllightcurves} 
\end{figure*}
\begin{figure*}
\includegraphics[width=165mm,clip]{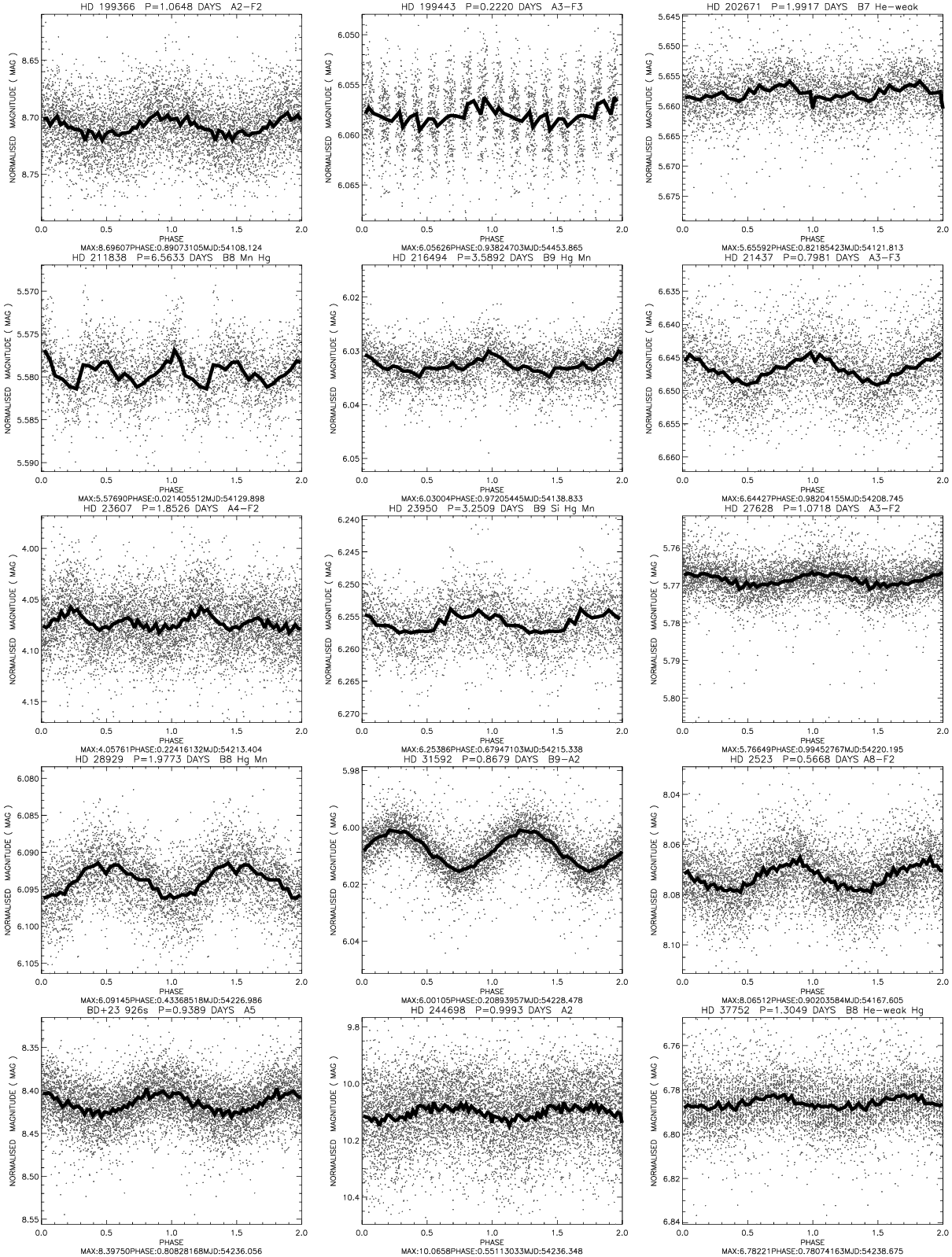}
\end{figure*}
\begin{figure*}
\includegraphics[width=165mm,clip]{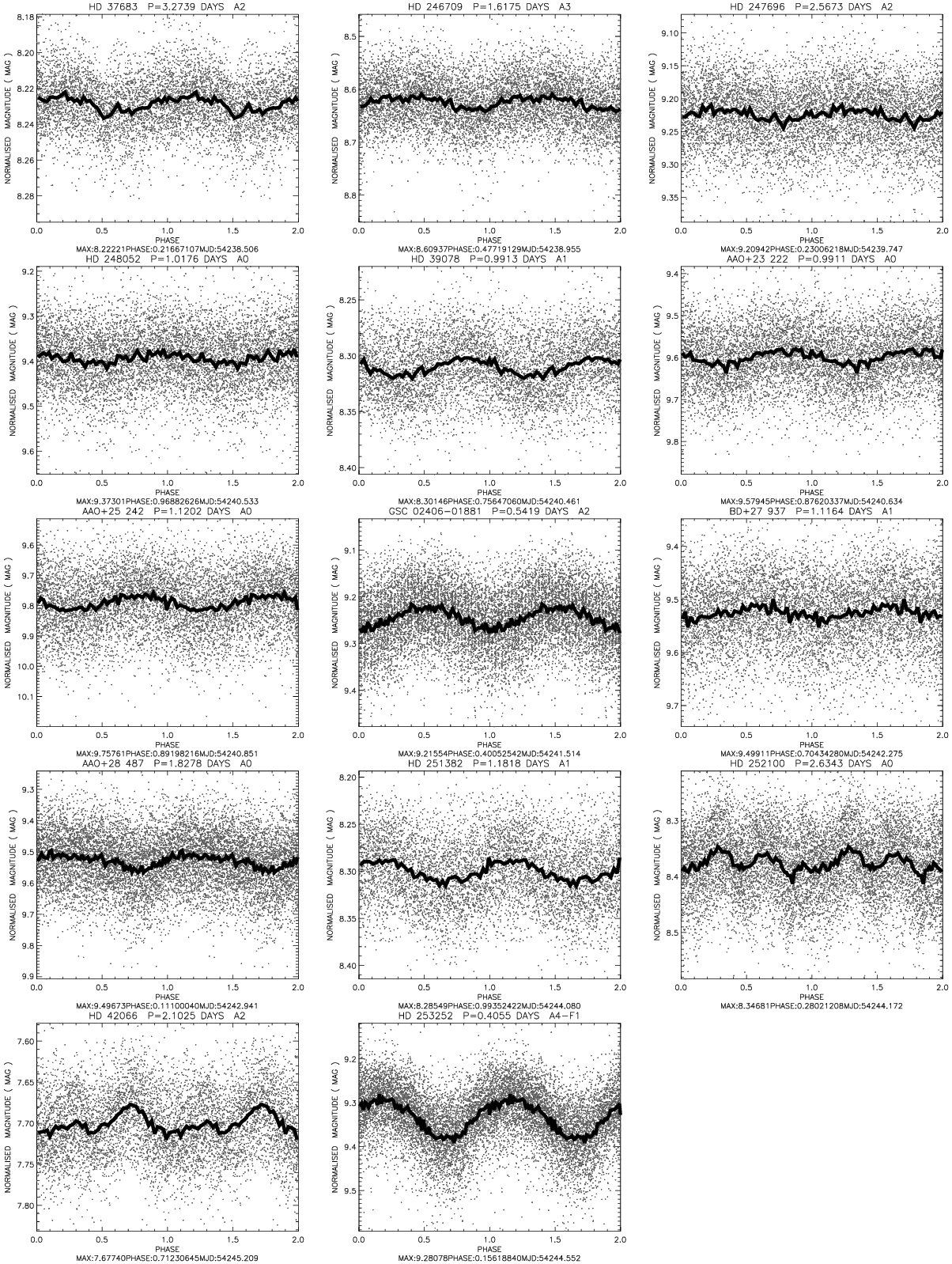}
\end{figure*}
\end{center} 

\bsp

\label{lastpage}

\end{document}